\documentclass[preprint,12pt]{elsarticle}

\usepackage{amsmath}
\usepackage{amssymb,color}
\newcommand{\dps}{\displaystyle}

\newcommand{\be}{\begin{equation}} 
\newcommand{\ee}{\end{equation}} 
\newcommand{\bea}{\begin{eqnarray}} 
\newcommand{\eea}{\end{eqnarray}}

\def\slash#1{#1\!\!\!\raise.15ex\hbox {/}}
\newcommand{\slD}{\,\raise.15ex\hbox{$/$}\kern-.27em\hbox{$\!\!\!D$}}
\newcommand{\slpartial}{\raise.15ex\hbox{$/$}\kern-.57em\hbox{$\partial$}}

\def\2int{\int_{0}^{1}  \!\!\! du \! \int_{0}^{1}  \!\!\! dv}

\def\e{\mbox{e}}

\def\Z{{\mathchoice {\hbox{$\sf\textstyle Z\kern-0.4em Z$}}
{\hbox{$\sf\textstyle Z\kern-0.4em Z$}}
{\hbox{$\sf\scriptstyle Z\kern-0.3em Z$}}
{\hbox{$\sf\scriptscriptstyle Z\kern-0.2em Z$}}}}

\def\no{\noindent}

\def\tanh{\rm tanh}
\def\half{\frac{1}{2}}

\def\mn{{\mu\nu}}

\def\tr{{\rm tr}\,}

\def\e{\,{\rm e}}
\def\b0{{\bf 0}}
\def\bear{\begin{eqnarray}}
\def\ear{\end{eqnarray}\noindent}
\def\bec{\blue\begin{equation}}
\def\eec{\end{equation}\black\noindent}
\def\bearc{\blue\begin{eqnarray}}
\def\earc{\end{eqnarray}\black\noindent}
\def\benn{\begin{enumerate}}
\def\enn{\end{enumerate}}

\def\intdp3{\int\frac{d^3p}{(2\pi)^3}}
\def\intdp4{\int\frac{d^4p}{(2\pi)^4}}
\def\kinb{\frac{\dot x^2}{4}}

\def\4piTD{{(4\pi T)}^{-{D\over 2}}}
\def\4piT4{{(4\pi T)}^{-2}}

\newcommand{\slG}{{{\dot G}\!\!\!\! \raise.15ex\hbox {/}}}

\def\dps{\displaystyle}

\begin{document}

\begin{frontmatter}



\title{
Low-energy limit of N-photon amplitudes\\ in a constant field
%\tnoteref{This article is registered under preprint number: arXiv:2105.08173 [hep-th]}
}


\author[label1]{Naser Ahmadiniaz}

\author[label1,label2]{Misha A. Lopez-Lopez}

\author[label3]{Christian Schubert}

%\affiliation[label1]{organization={Helmholtz-Zentrum Dresden-Rossendorf},
 %            addressline={Bautzner Landstra\ss e 400},
  %           city={Dresden},
   %          postcode={01328},
    %         country={Germany}}
     \address[label1]{Helmholtz-Zentrum Dresden-Rossendorf, Bautzner Landstraße 400, 01328 Dresden, Germany}
        
  %\affiliation[label2]{organization={Institut für Theoretische Physik, Technische Universität Dresden},%Department and Organization
         %   addressline={}, 
            %city={Dresden},
            %postcode={01062}, 
          %  state={},
            %country={Germany}}

   \address[label2]{Institut für Theoretische Physik, Technische Universität Dresden, 01062 Dresden, Germany}

%\affiliation[label3]{organization={Facultad de Ciencias Físico-Matemáticas, Universidad Michoacana de San Nicolás de Hidalgo},
 %            addressline={Avenida Francisco J. Mújica},
  %           city={Morelia},
   %          postcode={58060},
    %         state={Michoacán},
     %        country={Mexico}}

%\affiliation{organization={},%Department and Organization

   \address[label3]{Facultad de Ciencias Físico-Matemáticas, Universidad Michoacana de San Nicolás de Hidalgo, Avenida Francisco J. Mújica, 58060 Morelia, Michoacán, Mexico}

\begin{abstract}
While the QED photon amplitudes at full momentum so far have been calculated only up to the six-photon level, in the low-energy limit
there are explicit formulas for all helicity components even at the $N$-photon level, obtained by Martin et al. in 2002. 
Here we use the worldline formalism to extend that result to the $N$-photon amplitudes in a generic constant field.
For both scalar and spinor QED, we obtain compact representations for the low-energy limits of these amplitudes
involving only simple algebra and a single global proper-time integral with trigonometric integrand. 
% for which exact solutions of the Klein-Gordon and Dirac equations still exist. 
\end{abstract}

\end{frontmatter}


\section{Introduction}
\label{intro}

Although the QED photon amplitudes are prototypical for all gauge-boson correlators and have been studied for almost a century \cite{eulkoc,eulhei,weisskopf,aklapo,karneu50,karneu51,cotopi}
(for reviews, see \cite{liacza,scharnhorst,136}), for arbitrary momenta results exist only up to the six-point level \cite{bergui}. 
Only in the low-energy limit, where all photon energies are small compared to the electron mass, 
has it been feasible to derive explicit expressions for all helicity components even at the $N$-photon level \cite{56}. In the present letter, we generalize these formulas to the inclusion
of a generic constant external field. While in \cite{56} these amplitudes were derived from the
Euler-Heisenberg Lagrangian \cite{eulhei} and its scalar QED analogue, the Weisskopf Lagrangian \cite{weisskopf},
for the purpose of the constant-field generalization we find it more convenient to proceed along the lines of a later rederivation \cite{51} based on a direct amplitude calculation using the 
worldline formalism \cite{feynman:pr80,feynman:pr84,polyakov-book,berkosNPB,strassler1,5,41,Edwards:2019eby}. 
We will thus start with a short review of the calculation of general
photon amplitudes in that formalism, first in vacuum and then in the constant-field background.
We then proceed to the calculation of their low-energy limits, first for scalar and then for spinor QED. 

\section{Worldline representation of the QED $N$ - photon amplitudes} 

We start with a short summary of the computation of photon amplitudes in the worldline formalism (for details, see \cite{41,introWL}). 
In scalar QED, the formalism leads to the following path-integral representation of the one-loop $N$ - photon amplitude \cite{strassler1}, 
\bear
\Gamma_{\rm scal}(k_1,\varepsilon_1;\ldots ; k_N,\varepsilon_N) &=&
(-ie)^N 
\int_0^{\infty}
\frac{dT}{T}\,
\e^{-m^2T}
\int_{x(0)=x(T)}
Dx\,
{\rm e}^{-\int_0^T \! d\tau
\kinb}
\nonumber\\
&& \times
 V^{\gamma}_{\rm scal}[k_1,\varepsilon_1]\cdots V^{\gamma}_{\rm scal}[k_N,\varepsilon_N] \;.
 %\Bigl\vert_{{\rm lin}(\varepsilon_1,\ldots,\varepsilon_N)}
\label{10-Nphotonvertop}
\ear
Here $m$ is the mass of the loop scalar, the path integral runs over all closed loops of periodicity $T$ in (euclidean) spacetime, and each of the photons is represented by a
vertex operator 
\bear
V^{\gamma}_{\rm scal}[k,\varepsilon] = \int_0^Td\tau \, \varepsilon \cdot\dot x \e^{ik\cdot x}
=
\int_0^Td\tau \,\e^{ik\cdot x + \varepsilon \cdot\dot x}\Bigl\vert_{\varepsilon}\,,
\ear
where the polarisation vector $\varepsilon^\mu$ need not necessarily obey on-shell conditions. 
A formal gaussian integration of the path integral in \eqref{10-Nphotonvertop}
 leads to the following master formula for this amplitude
\cite{polyakov-book,berkosNPB,strassler1}:
\begin{eqnarray}
\Gamma_{\rm scal}(k_1,\varepsilon_1;\ldots;k_N,\varepsilon_N)
&=&
{(-ie)}^N
%{(2\pi )}^D\delta (\sum k_i)
{\dps\int_{0}^{\infty}}\frac{dT}{T}
{(4\pi T)}^{-\frac{D}{2}}
\e^{-m^2T}
\nonumber\\
&& \hspace{-150pt}
%\!\!\!\!\!\!\!
\times
\prod_{i=1}^N \int_0^T 
d\tau_i
\exp\biggl\lbrace\sum_{i,j=1}^N 
\Bigl\lbrack  \half G_{ij} k_i\cdot k_j
-i\dot G_{ij}\varepsilon_i\cdot k_j
+\half\ddot G_{ij}\varepsilon_i\cdot\varepsilon_j
\Bigr\rbrack\biggr\rbrace
\Bigl\vert_{\varepsilon_1\varepsilon_2\ldots \varepsilon_N}
\, .
%\nonumber\\
\label{scalarqedmaster}
\end{eqnarray}
\no
Here we have abbreviated $G_{ij}\equiv G(\tau_i,\tau_j)$ etc., where $G$ is the ``bosonic worldline Green's function'' defined by
\bear
G(\tau,\tau') \equiv \mid \tau-\tau'\mid 
-\frac{(\tau-\tau')^2}{T} \,,
\label{defG}
\ear
and a `dot' denotes a derivative acting on the first variable,
\begin{eqnarray}
\dot G(\tau,\tau') &=& {\rm sgn}(\tau - \tau')
- 2 \frac{(\tau - \tau')}{T}, \quad
\ddot G(\tau,\tau')
= 2 {\delta}(\tau - \tau')
- \frac{2}{T}
\, .
\label{GdGdd}
\end{eqnarray}
\noindent
%The factor ${(4\pi T)}^{-\frac{D}{2}}$
%represents the free Gaussian path integral
%determinant factor, and the 
%${(2\pi )}^D\delta (\sum k_i)$ factor is produced by the integration over the zero mode $x_0^{\mu}\equiv \frac{1}{T}\int_0^Td\tau\, x^\mu(\tau)$
%of the path integral.
The exponential must still be expanded and only the terms be retained that contain
each polarisation vector $\varepsilon_i$ linearly:
\bear 
\exp\bigl\lbrace 
\cdot
\bigr\rbrace 
\bigl\vert_{\varepsilon_1\varepsilon_2\ldots \varepsilon_N}
 \quad\equiv
 {(-i)}^N P_N(\dot G_{ij},\ddot G_{ij})
 \exp\biggl[\half \sum_{i,j=1}^N G_{ij}k_i\cdot
k_j \biggr] \,,
\label{defPN}
\ear
with certain polynomials $P_N$. 

The generalization to the spinor-loop case in the modern approach is done through the addition
of a Grassmann path integral, to be evaluated with the ``fermionic worldline Green's function'' $G_F(\tau,\tau') \equiv {\rm sgn}(\tau-\tau')$
\cite{fradkin,polyakov-book,berkosNPB,strassler1}. Moreover, these spin-induced terms can be conveniently
generated by the following ``Bern-Kosower loop-replacement" rule: after removing all second derivatives $\ddot G_{ij}$ from the polynomial $P_N$ by suitable
partial integrations, the integrand for the spinor loop case can be obtained by simultaneously replacing every ``$\tau$-cycle''  
$\dot G_{i_1i_2} 
\dot G_{i_2i_3} 
\cdots
\dot G_{i_ni_1}$
by the corresponding ``super $\tau$-cycle'' 
$\dot G_{i_1i_2} 
\dot G_{i_2i_3} 
\cdots
\dot G_{i_ni_1}
-
G_{Fi_1i_2}
G_{Fi_2i_3}
\cdots
G_{Fi_ni_1}
$
(and multiplying the whole amplitude by a factor of $-2$ for the difference in  statistics and number of degrees of freedom).  

Subsequently, it emerged that this whole procedure can be generalized to the inclusion
of a constant external field in a very economical way, namely by a modification of the path-integral determinant \cite{5} and
the introduction of generalized worldline Green's functions ${\cal G}_B,{\cal G}_F$ that take the external field into account 
\cite{shaisultanov,18}
\begin{eqnarray}
{\cal G}_{B}(\tau,\tau') &\equiv& \frac{T}{2{\cal Z}^2}
\biggl({{\cal Z}\over{{\rm sin}{\cal Z}}}
\,{\rm e}^{-i{\cal Z}\dot G(\tau,\tau')}
+i{\cal Z}\dot G(\tau,\tau') -1\biggr)\,,
\label{calGB}\\
{\cal G}_F(\tau,\tau') &\equiv&
G_F(\tau,\tau')
{{\rm e}^{-i{\cal Z}\dot G (\tau,\tau')}\over {\rm cos}{\cal Z}}\,,
\label{calGF}
\end{eqnarray}
\noindent
where ${\cal Z}_{\mu\nu} \equiv eF_{\mu\nu}T$.
These expressions for the constant-field Green's functions
should be understood as power series in the Lorentz matrix $\cal Z$.
The master formula for the photon amplitudes in vacuum \eqref{scalarqedmaster} can then be generalized to the
constant field case as follows \cite{shaisultanov,18},
\begin{eqnarray}
&&\Gamma_{\rm scal}
(k_1,\varepsilon_1;\ldots;k_N,\varepsilon_N;F)
=
{(-ie)}^N
%{(2\pi )}^D\delta^D (\sum k_i)
{\dps\int_{0}^{\infty}}{dT\over T}
{(4\pi T)}^{-{D\over 2}}
\e^{-m^2T}
{\rm det}^{{1\over 2}}
\biggl[
\frac{\cal Z}{{\rm sin}{\cal Z}}
\biggr]
\nonumber\\
&&\hspace{-20pt}\times
\prod_{i=1}^N \int_0^T \!\!\!
d\tau_i
\exp\biggl\lbrace\sum_{i,j=1}^N 
\Bigl\lbrack \half k_i\cdot {\cal G}_{Bij}\cdot  k_j
-i\varepsilon_i\cdot\dot{\cal G}_{Bij}\cdot k_j
+\half
\varepsilon_i\cdot\ddot {\cal G}_{Bij}\cdot\varepsilon_j
\Bigr\rbrack\biggr\rbrace
\Big\vert_{\varepsilon_1\varepsilon_2\cdots \varepsilon_N}\,. 
%\mid_{\rm multi-linear}\quad
\nonumber\\
\label{masterF}
\end{eqnarray}
\no
This representation has already been tested on the calculations of the tadpole \cite{112,113} and 
vacuum polarization amplitudes \cite{ditsha,40}, as well as on
magnetic photon splitting \cite{17,adler-book}, and found to be vastly more efficient than the traditional
methods based on second quantization. 

Similar master formulas have been derived for the $N$ - photon amplitudes in plane-wave backgrounds \cite{141} and in combined constant field -- plane-wave backgrounds \cite{154}. 

\section{Low-energy limit of the $N$ - photon amplitudes in vacuum}
\label{subsec:Low-energy limit of the N - photon amplitudes}

\begin{figure}[h]
   \centering
    \includegraphics[width=0.15\textwidth]{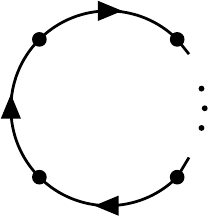}
 \caption{Worldline Feynman diagram representing an integrated bicycle factor.}
      \label{fig-srednicki}
      \end{figure}

The low-energy limit of the photon amplitudes is defined by the condition that all photon energies
be small compared to the mass of the loop scalar or fermion,  
\bear
\omega_i \ll m, \quad  i=1,\ldots, N \, .
\ear
This condition then justifies truncating all the vertex operators to their terms linear in the momentum. Noting 
that the leading, momentum-independent term in this expansion integrates to zero for a closed loop, and
adding a suitable total-derivative term, we can write the vertex operator of a low-energy photon as
\bear
V^{\gamma\, ({\rm LE})}_{\rm scal} [f] & = & 
\frac{i}{2} \int_0^Td\tau \, x(\tau)\cdot f \cdot \dot x(\tau)= \frac{i}{2}\int_0^T d\tau \e^{x(\tau)\cdot f\cdot\dot{x}(\tau)}\Big\vert_{f}\,,
\label{defVgammaLEfin}
\ear
where $f_{\mu\nu} = k_\mu \varepsilon_\nu - \varepsilon_\mu k_\nu$ is the photon field-strength tensor. 
The Wick contraction of a product of such objects produces products of ``Lorentz cycles'' 
\bear
Z_n(i_1i_2\ldots i_n)&\equiv&
\Bigl(\frac{1}{2}\Bigr)^{\delta_{n2}}
{\rm tr}
\Bigl(
\prod_{j=1}^n
f_{i_j}\Bigr) \,,
\label{defZn}
\ear\no
with coefficients that, by suitable partial integrations, can be written as integrals of the ``$\tau$ - cycles'' $\dot G_{i_1i_2}\dot G_{i_2i_3} \cdots \dot G_{i_ni_1}$ introduced above. 
The result can be further simplified by observing that, in the one-dimensional worldline theory, each of the resulting ``bicycle'' factors 
\bear
\int_0^T d\tau_{i_1} \cdots \int_0^Td\tau_{i_n} \dot G_{i_1i_2}\dot G_{i_2i_3} \cdots \dot G_{i_ni_1}\tr (f_{i_1}f_{i_2}\cdots f_{i_n})\,,
\label{bicycle}
\ear
can be identified with the one-loop $n$ - point Feynman diagram depicted in Fig. \ref{fig-srednicki}.

Products of such factors thus in the worldline theory correspond to 
disconnected diagrams, and by standard combinatorics can be reduced to the exponential of the
sum of all connected diagrams. In this way, and with a rescaling $\tau_i = Tu_i$, we arrive at 
\bear
\bigl\langle
V^{\gamma\, ({\rm LE})}_{\rm scal} [f_1] 
%V^{\gamma\, ({\rm LE})}_{\rm scal} [f_2] 
\cdots
V^{\gamma\, ({\rm LE})}_{\rm scal} [f_N] 
\bigr\rangle
=
(iT)^N 
\,\exp\biggl\lbrace \sum_{n=1}^{\infty}b_{2n}
\sum_{\lbrace i_1\ldots i_{2n}\rbrace}
Z^{\rm dist}_{2n}(\lbrace i_1i_2\ldots i_{2n}\rbrace)\biggr\rbrace
\bigg \vert_{f_1\ldots f_N}\,,
\nonumber\\
\label{WickLE}
\ear
where 
$Z^{\rm dist}_k(\lbrace i_1i_2\ldots i_k\rbrace)$
denotes the sum over all distinct
Lorentz cycles which can be formed with a given
subset of indices, e.g. 
$Z^{\rm dist}_4(\lbrace ijkl\rbrace) = Z_4(ijkl) + Z_4(ijlk) + Z_4(ikjl)$,
and $b_n$ denotes the basic ``cycle integral''
\bear
b_n &\equiv& \int_0^1 du_1du_2\ldots du_n\,
\dot G_{12}\dot G_{23}\cdots\dot G_{n1}  \;.
\label{cycleint} 
\ear
This integral can be expressed in terms of the Bernoulli numbers ${\cal B}_n$ \cite{18}:
\bear
b_n 
 =
\qquad\left\{ \begin{array}{r@{\quad\quad}l}
-2^n{{\cal B}_n\over n!}  & \qquad n{\rm \quad even}\,,\\
0 & \qquad n{\rm \quad odd}\,.\\
\end{array} \right.
\label{10-bn}
\ear
Eq. \eqref{WickLE} can be further simplified using the combinatorial fact that
\bear
{\rm tr}\Bigl[(f_1+\ldots +f_N)^n\Bigr]\bigg\vert_{\rm all\,\,different}
&=&
2n \sum_{\lbrace i_1\ldots i_n\rbrace} Z^{\rm dist}_n(\lbrace i_1i_2\ldots i_n\rbrace)\,,
\qquad
\label{F=F}
\ear
($n\ne 0$). Introducing $f_{\rm tot} \equiv \sum_{i=1}^N f_i$, 
using all this in \eqref{10-Nphotonvertop}
and eliminating the $T$-integral,
we arrive at the following formula for the low-energy limit of the one-loop
$N$-photon ($N\geq 4$) amplitude in scalar QED \cite{51}: 
\bear
\Gamma_{\rm scal}^{({\rm LE})}
(k_1,\varepsilon_1;\ldots ;k_N,\varepsilon_N)
&=&
\frac{e^N \Gamma(N-2)}{(4\pi)^2m^{2N-4}}
\,\exp\biggl\lbrace \sum_{n=1}^{\infty}\frac{b_{2n}}{4n} \tr (f_{\rm tot}^{2n})
\biggr\rbrace
\bigg \vert_{f_1\ldots f_N}\,,
\nonumber\\
\label{Nphotlowfin}
\ear
(here we have exempted the trivial case $N=2$ to be able to set $D=4$).

Further, the above-mentioned Bern-Kosower loop replacement rule allows us to generalise this result to spinor QED 
simply by replacing the cycle integral (\ref{10-bn}) by the ``super - cycle integral'' 
\bear
%b_n-f_n &\equiv& 
\int_0^1 \!\!\! du_1du_2\ldots du_n\,
\Bigl(\dot G_{12}\dot G_{23}\cdots\dot G_{n1} 
-
G_{F12}G_{F23}\cdots G_{Fn1}\Bigr)
=
(2-2^n)\,b_n \, .
\label{supercycleint}
\ear
The only other change is a global factor of $(-2)$ for statistics and degrees of freedom. Therefore \cite{51}
\bear
\Gamma_{\rm spin}^{({\rm LE})}
(k_1,\varepsilon_1;\ldots ;k_N,\varepsilon_N)
&=& (-2)
\frac{e^N \Gamma(N-2)}{(4\pi)^2m^{2N-4}}
\nonumber\\ &&\hspace{-30pt} \times
\,\exp\biggl\lbrace \sum_{n=1}^{\infty}(1-2^{2n-1})\frac{b_{2n}}{2n} \tr (f_{\rm tot}^{2n})
\biggr\rbrace
\bigg \vert_{f_1\ldots f_N}
\, .
\label{Nphotlowfinspin}
\ear
Formulas equivalent to \eqref{Nphotlowfin} and \eqref{Nphotlowfinspin} were derived in \cite{56} from the Euler-Heisenberg Lagrangian and its scalar QED equivalent,
the Weisskopf Lagrangian, however 
as already mentioned we have found the above procedure more suitable for our present purpose of generalization to the constant-field background. 

Note that in the above derivation on-shell conditions have not yet  been used. In \cite{56} it was further shown how to obtain explicit expressions for all the helicity components of the
on-shell amplitudes, using the spinor helicity formalism. And for the special case of the ``all plus'' (or ``all minus'') amplitudes this calculation was carried further to the two-loop level \cite{51},
taking advantage of the explicit expressions obtained in \cite{50} for the two-loop corrections to the Euler-Heisenberg and Weisskopf Lagrangians in the special case of a self-dual field. 

Two interesting aspects of the on-shell $N$-photon amplitudes in the low-energy limit (at any loop order) have emerged in that work. First, 
in the helicity basis the low-energy amplitudes obey a ``double Furry theorem'', that is, they are non-vanishing only if
the number of positive and negative helicity photons are separately even \cite{56}. 
Second, in the spinor-helicity formalism the full dependence of the low-energy amplitudes 
on the momenta and polarizations can, for any number of photons and 
any given helicity component, be absorbed into a single invariant, effectively reducing the
amplitude to a single number. This holds at any loop order, and made it feasible to use these amplitudes for a study of the asymptotic properties of the photon S-matrix \cite{81,111,121}. 

%%%%%%%%%%%%%%%%%%%%%%%%%
\section{Low-energy limit of the $N$ - photon amplitudes in a constant field: Scalar QED}

Let us now return to the scalar QED case, and show how to modify the procedure of the previous section to take an additional constant external field $F_\mn$ into account. 
Apart from the global determinant factor exhibited in \eqref{masterF}, the only change is that the Wick contraction of the $N$ low-energy vertex operators 
\eqref{defVgammaLEfin} now must be performed using the generalized Green's function \eqref{calGB}. One still arrives at the same representation in terms of bicycle factors
\eqref{bicycle}, only that now, since the function replacing $\dot G(\tau,\tau')$,
\begin{eqnarray}
\dot{\cal G}_B(\tau,\tau')
=
{i\over {\cal Z}}\biggl({{\cal Z}\over{{\rm sin}{\cal Z}}}
\,{\rm e}^{-i{\cal Z}\dot G(\tau,\tau')}-1\biggr)
\, ,
\label{derivcalGB}
\end{eqnarray}
is non-trivial in Lorentz space, the Lorentz and $\tau$ - cycle factors will not any more factorize, but rather combine to form a single Lorentz trace $\dot{\mathcal{G}}_B(i_1i_2\dots i_{n})$, defined as
\begin{eqnarray}
%\dot{\mathcal{G}}_B(i)&~\equiv ~&\frac{1}{2}\mbox{tr}(f_{i}\cdot\dot{\mathcal{G}}_{Bii})=\varepsilon_{i}\cdot\dot{\mathcal{G}}_{Bii}\cdot k_{i}\, ,\nonumber\\
%\dot{\mathcal{G}}_B(i_{1}i_{2})&~\equiv~&\frac{1}{2}\mbox{tr}(f_{i_1}\cdot\dot{\mathcal{G}}_{Bi_1 i_2}\cdot f_{i_2}\cdot\dot{\mathcal{G}}_{Bi_2 i_1})\, ,\nonumber\\
%&~=~&\varepsilon_{1}\cdot\dot{\mathcal{G}}_{B12}\cdot k_2\varepsilon_2 \cdot \dot{\mathcal{G}}_{B21}\cdot k_1-\varepsilon_1\cdot \dot{\mathcal{G}}_{B12}\cdot\varepsilon_2k_2\cdot\dot{\mathcal{G}}_{B21}\cdot k_1\nonumber\\
\dot{\mathcal{G}}_B(i_1i_2\dots i_{n})&~\equiv ~&
\Bigl(\half\Bigr)^{\delta_{n1}+\delta_{n2}}
\mbox{tr}(f_{i_1}\cdot\dot{\mathcal{G}}_{Bi_1i_2}\cdot f_{i_2}\cdot\dot{\mathcal{G}}_{Bi_2i_3}\cdots f_{i_n}\cdot\dot{\mathcal{G}}_{Bi_ni_1}).
%~~~(n\geq 3) \, . 
\nonumber\\
\label{12-defbicycle}
\end{eqnarray}
Thus in the presence of the constant field \eqref{WickLE} generalizes to
\bear
\bigl\langle
V^{\gamma\, ({\rm LE})}_{\rm scal} [f_1] 
%V^{\gamma\, ({\rm LE})}_{\rm scal} [f_2] 
\cdots
V^{\gamma\, ({\rm LE})}_{\rm scal} [f_N] 
\bigr\rangle_F
=
(iT)^N 
\!\exp\biggl\lbrace \sum_{n=1}^{\infty}
\sum_{\lbrace i_1\ldots i_{n}\rbrace}
%\int_0^1 du_{i_1} \cdots \int_0^1du_{i_n}
\prod_{k=1}^n \int_0^1 du_{i_k}
\dot{\mathcal{G}}_B^{\rm dist} 
(\lbrace i_1i_2\ldots i_{n}\rbrace)
\biggr\rbrace
\bigg \vert_{f_1\ldots f_N}
\nonumber\\
\label{WickLEF}
\ear
(note that $n$ is now running over all integers, not only over the even ones)
and the challenge is to compute the generalization of the cycle integral \eqref{cycleint},
\bear
I^{\rm cyc}_{\rm scal}(f_1,f_2,\ldots,f_n;F) \equiv \int_0^1 du_1 \cdots \int_0^1du_n 
\, \dot{\mathcal{G}}_B(12\dots n)
\, .
\label{defIscal}
\ear
This can be done in the following way. Restricting ourselves now to the case of a generic constant field (both Maxwell
invariants nonzero), and choosing a Lorentz frame where both the magnetic and the electric field point along the
$z$ axis, the (euclidean) field strength tensor takes the form\footnote{Our conventions follow \cite{41}.}
\begin{equation}
F =
\left(
\begin{array}{*{4}{c}}
0&B&0&0\\
-B&0&0&0\\
0&0&0&iE\\
0&0&-iE&0
\end{array}
\right) \;.
\label{app-gd-Fspecial}
\end{equation}
Defining $z_+ \equiv eBT$ and $z_- \equiv ieET$ and the matrices 
$g_+$, $g_-$, and $r_+$, $r_-$ by
\bear
g_+\equiv
\left(
\begin{array}{*{4}{c}}
1&0&0&0\\
0&1&0&0\\
0&0&0&0\\
0&0&0&0
\end{array}
\right),\qquad
g_-\equiv
\left(
\begin{array}{*{4}{c}}
0&0&0&0\\
0&0&0&0\\
0&0&1&0\\
0&0&0&1
\end{array}
\right),\nonumber\\
\label{app-gd-gmat}
\ear
\vspace{-30pt}
\begin{equation}
r_+ \equiv
\left(
\begin{array}{*{4}{c}}
0&1&0&0\\
-1&0&0&0\\
0&0&0&0\\
0&0&0&0
\end{array}
\right),\qquad
r_- \equiv
\left(
\begin{array}{*{4}{c}}
0&0&0&0\\
0&0&0&0\\
0&0&0&1\\
0&0&-1&0
\end{array}
\right),
%\label{app-gd-defrmat}
%\vspace{4mm}
\nonumber
\end{equation}
one can then show the factorization
\bear
{\rm det}^{{1\over 2}}
\biggl[
\frac{\cal Z}{{\rm sin}{\cal Z}}
\biggr]
=
\frac{z_+z_-}{{\rm sinh} z_+ {\rm \sinh} z_-}\,,
\ear
and the decomposition
\bear
\dot{\cal G}_{B}(\tau_i,\tau_j)
&=& \sum_{\alpha =\pm }
S_{Bij}(z_{\alpha})\,g_{\alpha}^{\mu\nu}
-i
\sum_{\alpha =\pm }
A_{Bij}(z_{\alpha})\,r_{\alpha}^{\mu\nu}\,,
\label{decomposecalG}
\ear
with coefficient functions
\bear
S_{Bij}(z) &=&
{\sinh(z\,\dot G_{ij})\over \sinh z} \,,
\label{defS}\\
A_{Bij}(z) &=&
{\cosh(z \,\dot G_{ij})\over 
\sinh z}-{1\over z} \;.
\label{app-defA}
\ear
For our present purpose, it is essential to note that those can be written in terms of the single function 
\bear
H^B_{ij} (z) \equiv 
\frac{e^{z \dot G_{ij}}}{\sinh z} - \frac{1}{z} 
\label{defHB}
\ear
as
\begin{eqnarray}
%\dot G_{ij} &=& H_{ij}(0) \\
S_{Bij}(z) &=& \half \Bigl\lbrack H^B_{ij}(z) + H^B_{ij}(-z) \Bigr\rbrack\,,
\\
A_{Bij}(z) &=& \half \Bigl\lbrack H^B_{ij}(z) - H^B_{ij}(-z) \Bigr\rbrack
\, .
%\dot{\cal G}_{Bij}
%&=&H_{ij}(0)\,{g_-}
%+ \frac{H_{ij} (z) + H_{ij}(-z)}{2} g_
%+ -A_{B12}(z) i{r_+}
\label{dotGbyH}
\ear
Thus, defining
\bear
\quad {\mathfrak g}_{\beta}^{\alpha}  \equiv \half (g_{\beta} - \alpha ir_{\beta}),
\ear
with $\alpha,\beta = \pm$,
we can rewrite \eqref{decomposecalG} as
\bear
\dot{\cal G}_{B}(\tau_i,\tau_j)
&=& \sum_{\alpha,\beta =\pm } H^B_{ij}(\alpha z_{\beta}) {\mathfrak g}_{\beta}^{\alpha}
\, .
\label{GBH}
\ear
The function $H^B$ has the following remarkable property of reproducing itself under folding, 
\bear
H_{ik}^{B(2)}(z,z') &\equiv &
\int_0^Td\tau_j H^B_{ij}(z) H^B_{jk}(z') = 
\frac{H^B_{ik}(z)}{z'-z} + \frac{H^B_{ik}(z')}{z-z'}\,,
\label{H2}\\
H_{il}^{B(3)}(z,z',z'') &\equiv &
\int_0^Td\tau_j \int_0^T d\tau_k H^B_{ij}(z) H^B_{jk}(z') H^B_{kl}(z'') 
\nonumber\\
&=&
\frac{H^B_{il}(z)}{(z'-z)(z''-z)}
+\frac{H^B_{il}(z')}{(z-z')(z''-z')}
+\frac{H^B_{il}(z'')}{(z-z'')(z'-z'')}\,,
\nonumber\\
&\vdots & \nonumber\\
H^{B(n)}_{i_1i_{n+1}}(z_1,\ldots,z_n) &=& \sum_{k=1}^n \frac{H^B_{i_1i_{n+1}}(z_k)}{\prod_{l \ne k} (z_l - z_k)}
\, .
\label{Hn}
\ear
Using this in \eqref{12-defbicycle}, \eqref{defIscal} we obtain
\bear
I^{\rm cyc}_{\rm scal}(f_1,\ldots,f_n;F) 
&=&
\Bigl(\half\Bigr)^{\delta_{n1}+\delta_{n2}}
\sum_{\alpha_1,\beta_1=\pm}
\cdots
\sum_{\alpha_n,\beta_n=\pm}
\tr (f_1 {\mathfrak g}_{\beta_1}^{\alpha_1}f_2\cdots  {\mathfrak g}_{\beta_n}^{\alpha_n})  
\nonumber\\&& \times
\sum_{k=1}^n \frac{H^B_{11}(\alpha_k z_{\beta_k})}{\prod_{l \ne k} (\alpha_l z_{\beta_l} - \alpha_k z_{\beta_k})}\,,
\label{Iscalfin}
\ear
where now only the coincidence limit of $H^B_{ij}(z)$ appears,
\bear
H^B_{ii}(z) = {\rm coth} \, z - \frac{1}{z}\,. 
\ear
Note that the product in the denominator can contain zero factors, but those are spurious and
cancelled by zeroes of the numerator (for an example, see the limit $z'\to z$ of the
right-hand side of \eqref{H2}).

Putting the pieces together, we arrive at the following generalization of the vacuum formula
\eqref{Nphotlowfin},
\bear
\Gamma_{\rm scal}^{({\rm LE})}
(k_1,\varepsilon_1;\ldots ;k_N,\varepsilon_N;F)
\!\!\!\!&=&\!\!\!\!
\frac{e^N}{(4\pi)^2}
\int_0^{\infty}\frac{dT}{T} T^{N-2} \e^{-m^2T}
\frac{z_+z_-}{{\rm sinh} z_+ {\rm \sinh} z_-}
\nonumber\\
&&\hspace{-40pt} \times
\exp\biggl\lbrace \sum_{n=1}^{\infty}\frac{1}{2n} 
I_{\rm scal}^{\rm cyc}(f_{\rm tot},\ldots,f_{\rm tot};F)
\biggr\rbrace
\bigg \vert_{f_1\ldots f_N}
\, .
\label{NphotlowfinF}
\ear

\section{Low-energy limit of the $N$ - photon amplitudes in a constant field: Spinor QED}

The transition to the spinor QED case can be done by the following adaption of the
Bern-Kosower replacement rule to the constant-field case \cite{18},
\bear
\dot{\mathcal{G}}_B(i_1i_2\dots i_{n})&\rightarrow&
\dot{\mathcal{G}}_B(i_1i_2\dots i_{n})-{\mathcal{G}}_F(i_1i_2\dots i_{n})\,,
\ear
with
\bear
{\mathcal{G}}_F(i_1i_2\dots i_{n})
&\equiv &
\Bigl(\half\Bigr)^{\delta_{n1}+\delta_{n2}}
\mbox{tr}(f_{i_1}\cdot{\mathcal{G}}_{Fi_1i_2}\cdot f_{i_2}\cdot{\mathcal{G}}_{Fi_2i_3}\cdots f_{i_n}\cdot{\mathcal{G}}_{Fi_ni_1})\,,
\nonumber\\
\ear
where $\mathcal{G}_F(\tau,\tau')$ was given in \eqref{calGF}. It permits a matrix decomposition
analogous to \eqref{GBH},
\bear
{\cal G}_{F}(\tau_i,\tau_j)
&=& \sum_{\alpha,\beta =\pm } H^F_{ij}(\alpha z_{\beta}) {\mathfrak g}_{\beta}^{\alpha}\,,
\label{GBHS}
\ear
now in terms of the function
\bear
H^F_{ij} (z) \equiv 
G_{Fij} \frac{e^{z \dot G_{ij}}}{\cosh z} 
\, .
\label{defHF}
\ear
Remarkably, this function turns out to obey exactly the same multiple-folding formula as $H^B$, eq. \eqref{Hn}. Thus without further ado we can generalize \eqref{Iscalfin} to
\bear
I^{\rm cyc}_{\rm spin}(f_1,\ldots,f_n;F) 
&\equiv&
\int_0^1 du_1 \cdots \int_0^1du_n \, 
\bigl( \dot{\mathcal{G}}_B(12\dots n) - {\mathcal{G}}_F(12\dots n)
\bigr)
\nonumber\\ 
&&\hspace{-100pt} =
\Bigl(\half\Bigr)^{\delta_{n1}+\delta_{n2}}
\sum_{\alpha_1,\beta_1=\pm}
\cdots
\sum_{\alpha_n,\beta_n=\pm}
\tr (f_1 {\mathfrak g}_{\beta_1}^{\alpha_1}f_2\cdots  {\mathfrak g}_{\beta_n}^{\alpha_n})  
\sum_{k=1}^n \frac{H^B_{11}(\alpha_k z_{\beta_k})-H^F_{11}(\alpha_k z_{\beta_k})}
{\prod_{l \ne k} (\alpha_l z_{\beta_l} - \alpha_k z_{\beta_k})}\,,
\nonumber\\
\label{Ispinfin}
\ear
which now involves only the function
\bear
H^B_{ii}(z) - H^F_{ii}(z) = {\rm coth} \, z -{\rm tanh} \, z - \frac{1}{z}\,. 
\ear
This brings us to our main result, the spinor QED generalization of the master formula
\eqref{NphotlowfinF} for the low-energy limit of the $N$ - photon amplitudes in the constant field:
\bear
\Gamma_{\rm spin}^{({\rm LE})}
(k_1,\varepsilon_1;\ldots ;k_N,\varepsilon_N;F)
\!\!\!\!&=&\!\!\!\!
-2
\frac{e^N}{(4\pi)^2}
\int_0^{\infty}\frac{dT}{T} T^{N-2} \e^{-m^2T}
\frac{z_+z_-}{{\rm \tanh} z_+ {\rm \tanh} z_-}
\nonumber\\
&&\hspace{-40pt} \times
\exp\biggl\lbrace \sum_{n=1}^{\infty}\frac{1}{2n} 
I_{\rm spin}^{\rm cyc}(f_{\rm tot},\ldots,f_{\rm tot};F)
\biggr\rbrace
\bigg \vert_{f_1\ldots f_N}\, .
\label{NphotlowfinFspin}
\ear

\section{Summary and Outlook}

To summarize, we have applied the worldline formalism to the calculation of the
low-energy limits of the $N$ - photon amplitudes in a generic constant field,
for both scalar and spinor QED.
Our final results for these amplitudes, \eqref{NphotlowfinF} and \eqref{NphotlowfinFspin}, 
reduce their explicit calculation to simple algebra and a single integral of Euler-Heisenberg type,
which could be easily automatized. They are still valid off-shell. 

In a separate publication \cite{alsst} we apply these formulas to the study of the on-shell four-photon amplitudes 
in a constant magnetic field. 
Further generalization from the closed-loop to the open-line case should be feasible along the lines of \cite{110,130,131}.

\bigskip

%\appendix
%
% \section{}
% \label{}



\end{document}